\documentclass{PoS}

\newcommand{\dhd}{{\textstyle d}
\lower.03ex\hbox{\kern-0.40em$^{\scriptstyle-}$}\kern-0.08em{}}  

\newcommand{\dbar}{{\textstyle \delta}
\lower.03ex\hbox{\kern-0.38em$^{\scriptstyle-}$}\kern-0.05em{}}

\newcommand{\bu}{{\bullet}}

\newcommand{\cald}{{\cal D}}  
\newcommand{\calf}{{\cal F}}

\newcommand{\calo}{{\cal O}}

\newcommand{\tilcaf}{\tilde{\cal F}} 
\newcommand{\ticalo}{\tilde{\cal O}} 
\newcommand{\ticalf}{\tilde{\cal F}}

\newcommand{\tilA}{\tilde{A}} 
 
\newcommand{\tiL}{\tilde{L}} 
\newcommand{\tilF}{\tilde{F}} 
 
\newcommand{\tilT}{\tilde{T}} 
\newcommand{\tilU}{\tilde{U}}

\title{Rapidity factorization and evolution of gluon TMDs}

\ShortTitle{Rapidity evolution of gluon TMDs}

\author{Ian Balitsky\\
        Theory Group, Jefferson Lab (JLAB), 12000 Jefferson Ave, Newport News, VA 23606,USA\\
        and\\
        Physics Dept.,  Old Dominion University, 4600 Elkhorn Ave, Norfolk VA 23529\\
        E-mail: \email{balitsky@jlab.org}}

\abstract{I discuss how the rapidity evolution of gluon transverse momentum dependent distribution 
changes from nonlinear evolution at small $x\ll 1$ to linear  evolution at moderate $x\sim 1$.}

\FullConference{QCD Evolution 2015 -QCDEV2015-\\
		26-30 May 2015\\
		Jefferson Lab (JLAB), Newport News, Virginia, USA}

\begin{document}

\section{Introduction}
A TMD factorization \cite{cs1,jimayuan,collinsbook} generalizes the usual concept of parton density by allowing PDFs to
depend on intrinsic transverse momenta in addition to the usual longitudinal momentum fraction variable. 
These transverse-momentum dependent parton distributions (also called unintegrated parton distributions) are widely used in the analysis of 
semi-inclusive processes like semi-inclusive deep inelastic scattering (SIDIS) 
or dijet production in hadron-hadron collisions (for a review, see Ref. \cite{collinsbook}). However, the analysis of TMD evolution in these cases is mostly
restricted to the evolution of quark TMDs, whereas at high collider energies the majority of produced particles
will be small-$x$ gluons. In this case one has to understand the transition between non-linear dynamics at small $x$ and
presumably linear evolution of gluon TMDs at intermediate $x$.

 In this presentation I discuss the connection between rapidity evolution of gluon TMD at low $x_B$
 and at moderate $x_B\sim 1$.  (The discussion is based on the paper \cite{BT1}).
We assume $k_\perp^2\geq $ few GeV$^2$ so that we can use perturbative QCD, but otherwise $k_\perp$ 
is arbitrary and can be of order of $s$ as in the DGLAP evolution. In this kinematic region we will vary Bjorken $x_B$ and look how non-linear evolution at small $x$ transforms
 into linear evolution at moderate $x_B$.  It should be noted that at least at moderate $x_B$ gluon TMDs mix with the quark ones. 
Here I disregard this mixing leaving the discussion of full matrix for future publications. 
   
It should be emphasized that here I consider gluon TMDs with Wilson links going to $+\infty$ in the longitudinal direction relevant for 
SIDIS \cite{collins1}.  Note that in the leading order SIDIS is determined solely by quark TMDs but beyond that the gluon TMDs 
should be taken into account, especially for the description of various processes at future EIC collider.
 
 The presentation is organized as follows. In Sec. 2 I remind the general logic of rapidity factorization and rapidity evolution. 
 In Sec. 3 I describe the evolution equation of gluon TMD in the light-cone limit.
 In Sec. 4 I present the Lipatov vertex of the gluon production by the $\calf^a_i$  operator and the so-called virtual corrections. 
 The final TMD evolution 
 equation for all $x_B$ and  transverse momenta is discussed in Sec. 5 while Sec. 6 contains conclusions and outlook. 

\section{Rapidity factorization and evolution \label{sec2}}

It is convenient to define the field-strength operator with attached light-like Wilson line: 
\begin{eqnarray}
&&\hspace{-0mm} 
\calf^{a\eta}_{i}( x_B,z_\perp)~\equiv~{2\over s}
\!\int\! dz_\ast ~e^{i x_B z_\ast} 
\big([\infty,z_\ast]_z^{am}gF^m_{\bu i}(z_\ast,z_\perp))^\eta
\label{kalf}
\end{eqnarray}
where the index $\eta$ denotes the rapidity cutoff (\ref{cutoff}) for all gluon fields in this operator:
\begin{eqnarray}
&&\hspace{-0mm} 
A^\eta_\mu(x)~=~\int\!{d^4 k\over 16\pi^4} ~\theta(e^\eta-|\alpha|)e^{-ik\cdot x} A_\mu(k)
\label{cutoff}
\end{eqnarray}
The  Sudakov variable $\alpha$ is defined as usual,  $k=\alpha p_1+\beta p_2+k_\perp$.
We define the light-like vectors $p_1$ and $p_2$ such that $p_1=n$  and $p_2=p-{m^2\over s}n$, where $p$ is the momentum of the target particle of mass $m$. 
We use metric $g^{\mu\nu}~=~(1,-1,-1,-1)$ so $p\cdot q~=~(\alpha_p\beta_q+\alpha_q\beta_p){s\over 2}-(p,q)_\perp$. For the coordinates we use 
the notations $x_\bu\equiv x_\mu p_1^\mu$ and $x_\ast\equiv x_\mu p_2^\mu$ related to the light-cone coordinates by $x_\ast=\sqrt{s\over 2}x_+$ and $x_\bu=\sqrt{s\over 2}x_-$.

Hereafter we use the notation 
$[\infty, z_\ast]_z\equiv[\infty_\ast p_1+z_\perp, {2\over s}z_\ast p_1+z_\perp]$  where  
$[x,y]$ stands for the straight-line gauge link connecting points $x$ and $y$. 
Our convention is that the Latin Lorentz indices always correspond to transverse coordinates while Greek Lorentz indices are four-dimensional.

Similarly, we define 
\begin{eqnarray}
&&\hspace{-0mm} 
\tilcaf^{a\eta}_i( x_B,z_\perp)~\equiv~{2\over s}
\!\int\! dz_\ast ~e^{-i x_B z_\ast} 
g\big(\tilF^m_{\bu i}(z_\ast,z_\perp)[z_\ast,\infty]_z^{ma}\big)^\eta
\label{tilkaf}
\end{eqnarray}
in the complex-conjugate part of the amplitude.

In this notations the unintegrated gluon TMD $\cald( x_B,z_\perp,\eta)$ can be represented as 
\begin{eqnarray}
&&\hspace{-0mm} 
\langle p|\tilcaf^{a\eta}_i( x_B,z_\perp)\calf^{ai\eta}( x_B,0_\perp)|p+\xi p_2\rangle 
\equiv\sum_X\langle p|\tilcaf^{a\eta}_i( x_B,z_\perp)|X\rangle\langle X|\calf^{ai\eta}( x_B,0_\perp)|p+\xi p_2\rangle 
\nonumber\\
&&\hspace{-0mm} 
=~-4\pi^2\delta(\xi) x_Bg^2\cald( x_B,z_\perp,\eta)
\label{TMD}
\end{eqnarray}
Hereafter we use a short-hand notation
\begin{eqnarray}
&&\hspace{-0mm} 
\langle p|\ticalo_1...\ticalo_m\calo_1...\calo_n |p'\rangle
\equiv~\sum_X\langle p| \tilT\{ \ticalo_1...\ticalo_m\}|X\rangle\langle X|T\{\calo_1...\calo_n\}|p'\rangle
\label{ourop}
\end{eqnarray}
where tilde on the operators in the l.h.s. of this formula stands as a reminder that they should be inverse time ordered
as indicated by inverse-time ordering $\tilT$ in the r.h.s. of the above equation.

As discussed e.g. in Ref. \cite{keld},  such martix element can be represented by a double functional integral
\begin{eqnarray}
&&\hspace{-0mm} 
\langle \ticalo_1...\ticalo_m\calo_1...\calo_n \rangle
\nonumber\\
&&\hspace{-0mm} 
=~\int\! D\tilA D\tilde{\bar\psi}D\tilde{\psi}~e^{-iS_{\rm QCD}(\tilA,\tilde{\psi})}\!\int\! DA D\bar{\psi} D\psi ~e^{iS_{\rm QCD}(A,\psi)} \ticalo_1...\ticalo_m\calo_1...\calo_n
\label{funtegral}
\end{eqnarray}
with the boundary condition $\tilA(\vec{x},t=\infty)=A(\vec{x},t=\infty)$ (and similarly for quark fields) reflecting the sum over all intermediate states $X$. 
Due to this condition, the matrix element (\ref{TMD}) can be made gauge-invariant by connecting the endpoints of Wilson lines at infinity with the gauge link 
\begin{eqnarray}
&&\hspace{-0mm} 
\langle p|\tilcaf^a_i( x_B,x_\perp)\calf^{ai}( x_B',y_\perp)|p'\rangle~
\nonumber\\
&&\hspace{-0mm} 
\rightarrow~\langle p|\tilcaf^a_i( x_B,x_\perp)[x_\perp+\infty p_1,y_\perp+\infty p_1]\calf^{ai}(\beta'_B,y_\perp)|p'\rangle
\label{inftylink}
\end{eqnarray}
This gauge link is important if we use the light-like gauge $p_1^\mu A_\mu=0$ for calculations \cite{bejuan}, but in all other gauges it can be neglected.
We will not write it down explicitly but will always assume it in our formulas.

We will study the rapidity evolution of the operator 
\begin{equation}
\tilcaf^{a\eta}_i( x_B,x_\perp)\calf^{a\eta}_j( x_B,y_\perp)
\label{operator}
\end{equation}

In the spirit of rapidity factorization, in order to find the evolution of the operator (\ref{operator})
with respect to rapidity cutoff $\eta$ 
(see Eq. (\ref{cutoff})) one should integrate in the matrix element  (\ref{TMD}) over gluons and quarks with rapidities $\eta>Y>\eta'$ and temporarily ``freeze'' fields with $Y<\eta'$ to be integrated over later. (For a review, see Refs. \cite{mobzor,nlolecture}.) In this case, we obtain functional integral of Eq. (\ref{funtegral}) type over fields with $\eta>Y>\eta'$ in the ``external'' fields with $Y<\eta'$. 
In terms of Sudakov variables we integrate over gluons with $\alpha$ between $\sigma=e^\eta$ and $\sigma'=e^{\eta'}$ and, in the leading order, only
the diagrams with gluon emissions are relevant - the quark diagrams will enter as loops at the next-to-leading (NLO) level.

To make connections with parton model we will have in mind the frame where target's velocity is large and call the small $\alpha$ fields by the name ``fast fields'' and large $\alpha$ fields by 
``slow'' fields.  As discussed in Ref. \cite{npb96},  the interaction of ``slow'' gluons of large $\alpha$ with ``fast'' fields of 
small $\alpha$  is described by eikonal  gauge factors and the integration over slow fields  results in  
Feynman diagrams in the background of fast fields which form a thin shock wave due to Lorentz contraction.
 However,  in Ref. \cite{npb96} (as well as in all small-$x$ literature)
it was assumed that the characteristic transverse momenta of fast and slow fields are of the same order of
magnitude. For our present purposes we need to relax this condition and consider cases where the 
transverse momenta of fast and slow fields do differ. In this case, we need to rethink the shock-wave approach.

Let us figure out how the relative longitudinal size of fast and slow fields depends on their transverse momenta.
 The typical longitudinal size of fast 
fields is $\sigma_\ast\sim {\sigma' s\over l_\perp^2}$ where $l_\perp$ is the characteristic 
scale of transverse momenta of fast fields. The typical distances  traveled by slow gluons are 
$\sim{\sigma s\over k_\perp^2}$ where $k_\perp$ is the characteristic 
scale of transverse momenta of slow fields. Effectively, the 
large-$\alpha$ gluons propagate in the external field of the small-$\alpha$ shock wave, except the case
$l_\perp^2\ll k_\perp^2$ which should be treated separately since the ``shock wave'' is not necessarily thin in this case.  
Fortunately, when $l_\perp^2\ll k_\perp^2$ one can use  
the light-cone expansion of slow fields and leave at the leading order only the light-ray operators of the leading twist.
We will use the combination of shock-wave and light-cone expansions and write the interpolating formulas 
which describe the leading-order contributions in both cases. 

\section{Evolution kernel in the light-cone limit}

As we discussed above, we will obtain the evolution kernel in two separate cases: the ``shock wave'' case when  the characteristic transverse momenta 
of the background gluon (or quark) fields $l_\perp$ are of the order of typical momentum of emitted gluon $k_\perp$ and  
the ``light cone'' case when $l_\perp^2\ll k_\perp^2$.  It is convenient to start with the light-cone situation and consider 
 the one-loop evolution of the operator $\tilcaf^{a\eta}_i( x_B,x_\perp)\calf^{ai\eta}( x_B,y_\perp)$ 
 in the case when the background fields are soft so we can use the 
 expansion of propagators in external fields near the light cone \cite{bbr}. 
 
\begin{figure}[htb]
\begin{center}
\includegraphics[width=104mm]{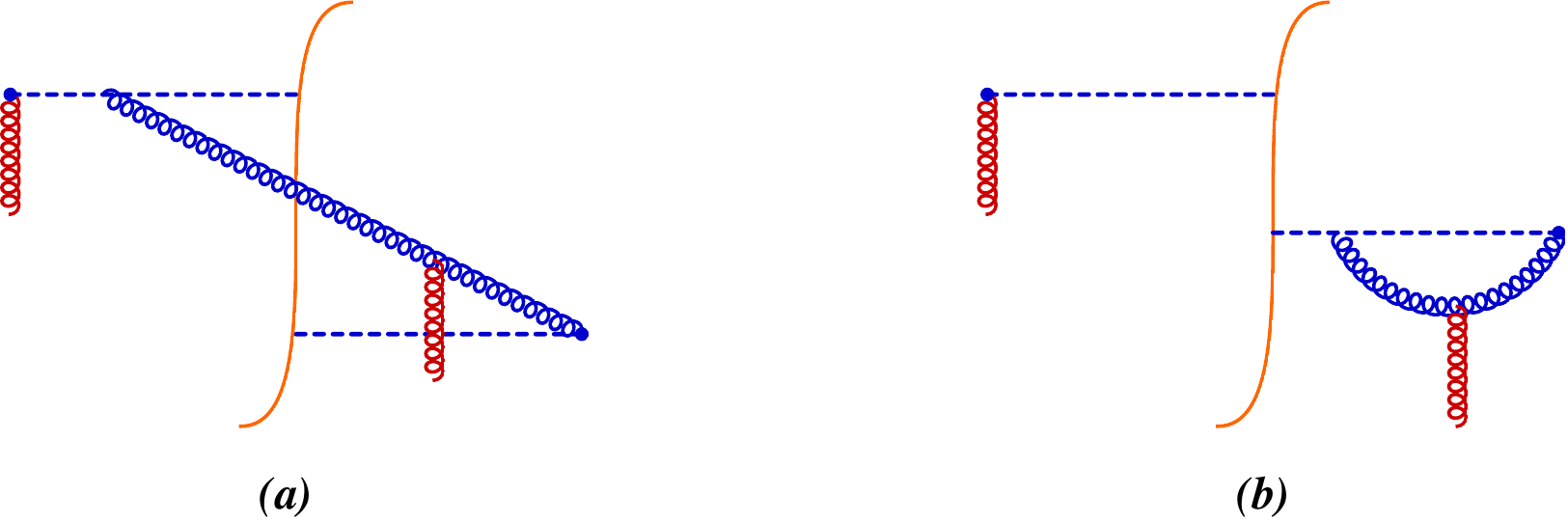}
\end{center}
\caption{Typical diagrams for production (a) and virtual (b) contributions to the evolution kernel. The dashed lines denote gauge links.\label{fig:1}}
\end{figure}

 In the leading order there is only one ``quantum'' gluon and we get the typical diagrams of Fig. \ref{fig:1} type. 
One sees that the evolution kernel consist of two parts: ``real'' part with the emission of a real gluon and a ``virtual'' part without such emission. 
The ``real'' production part of the kernel can be obtained as a square of a Lipatov vertex - the amplitude of the emission of a real gluon by the Wilson-line operator $\calf^a_i$:  
\begin{eqnarray}
&&\hspace{-1mm}
\langle \tilcaf_i^a( x_B,x_\perp) \calf_j^a( x_B,y_\perp)\rangle^{\ln\sigma}~
=~-\!\int_{\sigma'}^{\sigma}\!{\dhd\alpha\over 2\alpha}\!\int\!\dhd^2k_\perp
\langle \tiL_{ab}^{\mu i}(k,y_\perp, x_B)L^{ab}_{\mu i}(k,y_\perp, x_B)\rangle ^{\ln\sigma'}
\label{3.1}
\end{eqnarray}
where the Lipatov vertices of gluon emission are defined as  
\begin{eqnarray}
&&\hspace{-1mm}
L^{ab}_{\mu i}(k,y_\perp, x_B)~=~i\lim_{k^2\rightarrow 0}k^2\langle A^a_\mu(k)\calf^b_i( x_B,y_\perp)\rangle
\nonumber\\
&&\hspace{-1mm}
\tiL^{ba}_{i\mu}(k,x_\perp, x_B)
=-i\!\lim_{k^2\rightarrow 0}k^2\langle \tilcaf^{b}_i( x_B,x_\perp)\tilA_\mu^{a}(k) \rangle
\label{3.2}
\end{eqnarray}
($\tiL$ is a complex-conjugate vertex made of $\tilA$ fields).
Hereafter we use the space-saving notation $\dhd^np\equiv {d^np\over(2\pi)^n}$.
\footnote{To simplify our notations, we will often omit 
label $\eta$ for the rapidity cutoff (\ref{cutoff}) but it will be always assumed when not displayed.}

\begin{figure}[htb]
\begin{center}
\includegraphics[width=114mm]{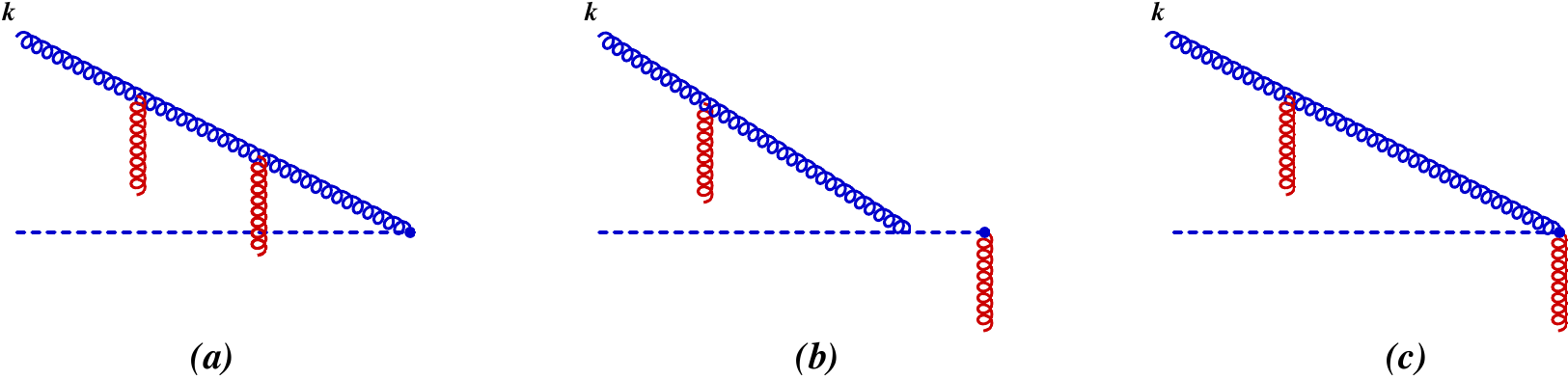}
\end{center}
\caption{Lipatov vertex of gluon emission near the light cone. \label{fig:2}}
\end{figure}

The three corresponding diagrams are shown in Fig. \ref{fig:2}. 
One obtains (in the light-like gauge $p_2^\mu A_\mu=0$)
\begin{eqnarray}
&&\hspace{-3mm}
L^{ab}_{\mu i}(k,y_\perp, x_B)^{\rm light-like}~=~2ge^{-i(k,y)_\perp}
\label{3.8}\\
&&\hspace{-3mm}
\times~\Big[
{k^\perp_\mu\delta_i^l\over k_\perp^2}
-{\delta_\mu^lk_i+\delta_i^lk^\perp_\mu-g_{\mu i}k^l\over \alpha x_Bs+k_\perp^2}-{k_\perp^2g_{\mu i}k^l+2k^\perp_\mu k_ik^l\over(\alpha x_Bs+k_\perp^2)^2}\Big]
\calf_l^{ab}( x_B+{k_\perp^2\over\alpha s},y_\perp)~+~O(p_{2\mu})
\nonumber
\end{eqnarray}
We do not write down the terms $\sim p_{2\mu}$ since they do not contribute to the production kernel ($\sim$ square of the expression in the r.h.s. of Eq. (\ref{3.8})).

The product of Lipatov vertex (\ref{3.8}) and the complex conjugate vertex $L^{ab}_{\mu i}(k,y_\perp, x_B)$  integrated according to Eq. (\ref{3.1}) gives 
the production part of the evolution kernel in the light-cone limit. 
To get the full kernel, we need to add the virtual contribution coming from diagrams of Fig. \ref{fig:1}b type  which has the form
\begin{eqnarray}
&&\hspace{-1mm}
\langle \tilcaf_i^a( x_B, x_\perp) \calf_j^a( x_B,y_\perp)\rangle_{\rm virt}~
\nonumber\\
&&\hspace{-1mm}
=~-2g^2N_c\tilcaf_i^a( x_B,x_\perp) \calf_j^a( x_B,y_\perp)\!\int_0^\infty\!{\dhd\alpha\over \alpha}
\!\int\! \dhd^2p_\perp {\alpha x_Bs\over p_\perp^2(\alpha x_Bs+p_\perp^2)}
\label{3.19}
\end{eqnarray}
where we used  Schwinger's notations
\begin{equation}
(x_\perp|f(p_\perp)|y_\perp)~\equiv~ \int\! \dhd^2p_\perp~e^{i(p,x-y)_\perp}f(p), ~~~~~(x_\perp|p_\perp)~=~e^{i(p,x)_\perp}
\label{schwingerperp}
\end{equation}

Note that with our rapidity cutoff in $\alpha$ (Eq. (\ref{cutoff})) the contribution (\ref{3.19}) coming from the diagram 
in Fig. \ref{fig:1}b is UV finite. 

Summing the product of Lipatov vertices (\ref{3.8}) 
the virtual correction (\ref{3.19}) we obtain the one-loop evolution kernel in the light-cone approximation
\begin{eqnarray}
&&\hspace{-1mm}
{d\over d\ln\sigma}\langle p|\ticalf_i^a( x_B,x_\perp)\calf_j^a( x_B,y_\perp)|p\rangle~
\label{3.21}\\
&&\hspace{-1mm}
=~{g^2N_c\over\pi}\!\int\!\dhd^2k_\perp~\Big\{e^{i(k,x-y)_\perp}
\langle p|\ticalf_k^a\big( x_B+{k_\perp^2\over\sigma s},x_\perp\big)\calf_l^a\big( x_B+{k_\perp^2\over\sigma s},y_\perp\big)
|p\rangle
\nonumber\\
&&\hspace{-1mm}
\times~\Big[{\delta_i^k\delta_j^l\over k_\perp^2}-{2\delta_i^k\delta_j^l\over\sigma x_Bs+k_\perp^2}
+~{k_\perp^2\delta_i^k\delta_j^l+\delta_j^kk_ik^l+\delta_i^lk_jk^k-\delta_j^lk_ik^k-\delta_i^kk_jk^l-g^{kl}k_ik_j-g_{ij}k^kk^l
\over (\sigma x_Bs+k_\perp^2)^2}
\nonumber\\
&&\hspace{2mm}
+~k_\perp^2{2g_{ij}k^kk^l+\delta_i^kk_jk^l+\delta_j^lk_ik^k-\delta_j^kk_ik^l-\delta_i^lk_jk^k\over (\sigma x_Bs+k_\perp^2)^3}
-{k_\perp^4g_{ij}k^kk^l\over  (\sigma x_Bs+k_\perp^2)^4}\Big]
\theta\big(1- x_B-{k_\perp^2\over \sigma s}\big)
\nonumber\\
&&\hspace{58mm}
-~ {\sigma x_Bs\over k_\perp^2(\sigma x_Bs+k_\perp^2)} 
\langle p|\ticalf_i^a( x_B,x_\perp)\calf_j^a( x_B,y_\perp)|p\rangle\Big\}
\nonumber
\end{eqnarray}
where $\theta\big(1- x_B-{k_\perp^2\over \sigma s}\big)$ is a kinematical restriction that the sum of $ x_B$ and the fraction carried by 
emitted gluon $ {k_\perp^2\over\alpha s}$ should be less than one (there is obviously no restriction on $k_\perp$ in the virtual diagram).

\section{Evolution kernel in the general case}

In this section we will find the leading-order rapidity evolution of gluon operator (\ref{operator}) 
with the rapidity cutoff $Y<\eta=\ln\sigma$ for all emitted gluons. As we mentioned in Sect. 2, 
in order to find the evolution kernel we need to integrate over slow gluons  
with $\sigma>\alpha>\sigma'$ and temporarily freeze fast fields with $\alpha<\sigma'$ to be integrated over later. 
To this end we need the one-loop diagrams in the fast background fields with arbitrary transverse momenta. 
In the previous section we have found the evolution kernel in background fields with  transverse momenta 
$l_\perp\ll p_\perp$ where $p_\perp$ is a characteristic momentum of our quantum slow fields. In this section
at first we will find the Lipatov vertex and virtual correction for the case $l_\perp\sim p_\perp $ and then write down 
general formulas which are correct in the whole region of the transverse momentum. 

The key observation is that for transverse momenta of quantum and background field of the same order we can use the shock-wave approximation developed for small-$x$ physics.  To find the evolution kernel we consider 
the operator (\ref{operator}) in the background of external field $A_\bu(x_\ast,x_\perp)$ (the absence of $x_\bu$ in the argument corresponds to $\alpha=0$). Moreover, we assume that the background field $A_\bu(x_\ast,x_\perp)$ has a narrow
support and vanishes outside the $[-\sigma_\ast,\sigma_\ast]$ interval. This is obviously not the most general form of the external field, 
but it turns out that after obtaining the kernel of the evolution equation it is easy to restore the result for any background field by 
insertion of gauge links at $\pm\infty p_1$, see the discussion after Eq. (\ref{6.1}).

Since the typical $\beta$'s of the external field are $\beta_{\rm fast}\sim {l_\perp^2\over\alpha_{\rm fast}s}$ the support of the shock wave $\sigma_\ast$ is of order of ${1\over \beta_{\rm fast}}\sim{\sigma's\over l_\perp^2}$. This is to be compared to the typical scale of slow fields ${1\over\beta_{\rm slow}}\sim {\alpha s\over p_\perp^2}\gg \sigma_\ast$ 
so we see that the fast background field can be approximated by a narrow shock wave.
In the ``pure'' low-x case $ x_B=0$ one can
assume that the support of this shock wave is infinitely narrow. As we shall see below, in our case of arbitrary $ x_B$ we need to look inside the shock wave so we will separate all integrals over longitudinal distances $z_\ast$ in 
parts ``inside the shock wave''
$|z_\ast|<\sigma_\ast$ and ``outside the shock wave" $|z_\ast|>\sigma_\ast$, calculate them separately and
check that the sum of ``inside'' and ``outside'' contributions does not depend on $\sigma_\ast$ with our accuracy.

\begin{figure}[htb]
\begin{center}
\includegraphics[width=104mm]{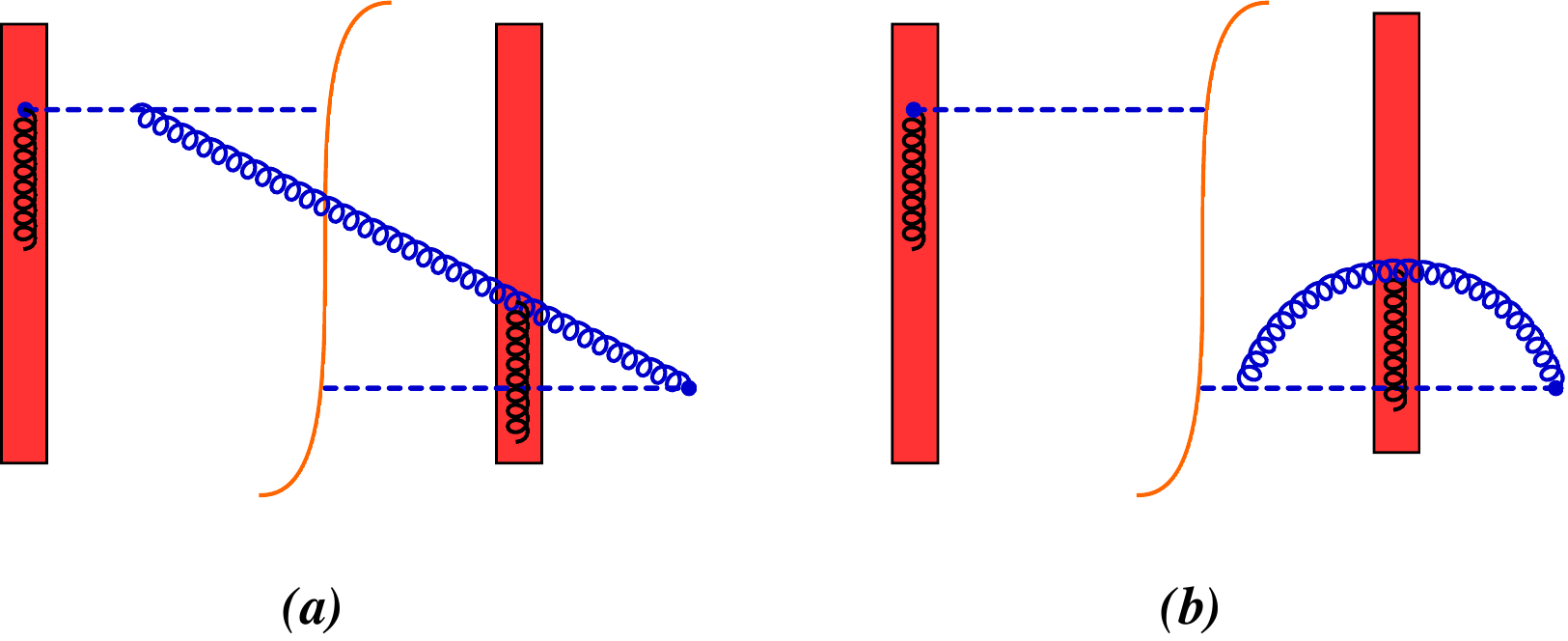}
\end{center}
\caption{Typical diagrams for production (a) and virtual (b) contributions to the evolution kernel. The shaded area denotes shock
wave of background fast fields. \label{fig:3}}
\end{figure}

In the leading order there is only one extra gluon and we get the typical diagrams of Fig. \ref{fig:3} type. 
The production part of the kernel can be obtained as a square of a Lipatov vertex - the amplitude of the emission of a real gluon by the operator $\calf^a_i$ (see Eq. (\ref{3.1}))
\begin{eqnarray}
&&\hspace{-8mm}
\langle \tilcaf^{a}_i( x_B, x_\perp) \calf_j^a( x_B, y_\perp)\rangle^{\ln\sigma}~
=~-\!\int_{\sigma'}^{\sigma}\!{\dhd\alpha\over 2\alpha}\dhd^2k_\perp
~\big(\tiL^{ba;\mu}_i(k,x_\perp, x_B)
L^{ab}_{\mu j}(k,y_\perp, x_B)\big)^{\ln\sigma'}
\label{4.1}
\end{eqnarray}
where the Lipatov vertices of gluon emission are defined in Eq. (\ref{3.2}) . Hereafter $\langle\calo\rangle$ means the average of operator $\calo$ in the shock-wave background.

\begin{figure}[htb]
\begin{center}
\includegraphics[width=141mm]{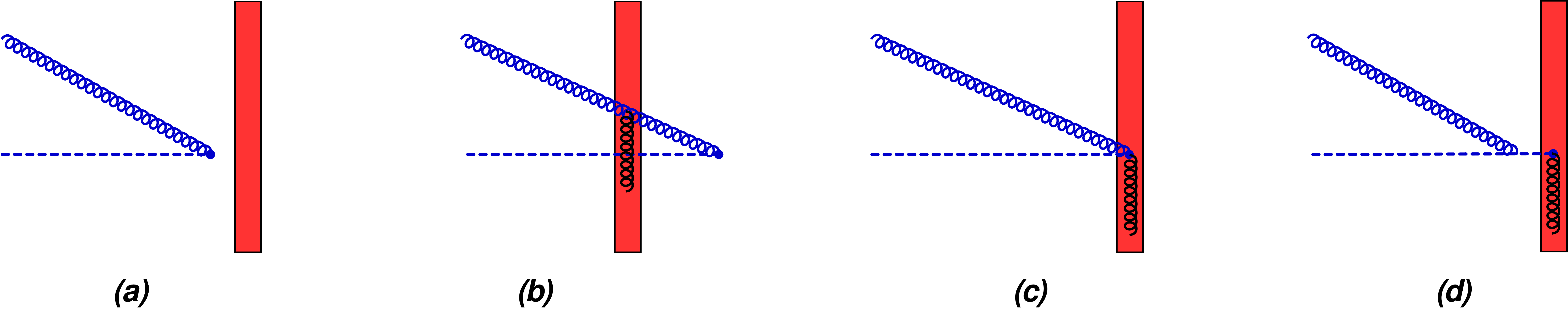}
\end{center}
\caption{Lipatov vertex of gluon emission in a shock-wave background. \label{fig:4}}
\end{figure}

As we discussed above, we calculate the diagrams in the background of a shock wave of 
width $\sim{\sigma' s\over l_\perp^2}$ 
where $l_\perp$ is the characteristic transverse momentum of the external shock-wave field. Note that the factor in the exponent
in the definition of $\calf( x_B)$ is $\sim  x_B{\sigma' s\over l_\perp^2}$ which is not necessarily small at various $ x_B$ and
$l_\perp^2$ and therefore we need to take into account the diagram in Fig. \ref{fig:4}c with emission point inside the shock wave. 
We do this in a following way: we assume that all of the shock wave is contained within $\sigma_\ast>z_\ast>-\sigma_\ast$,
calculate diagrams in  Fig. \ref{fig:4}a-d and check that the dependence on $\sigma_\ast$ cancels in the
final result for the sum of these diagrams. 
The result of the calculation is \cite{BT1}
\begin{eqnarray}
&&\hspace{-1mm}
L^{ab}_{\mu i}(k,y_\perp, x_B)^{\rm light-like}   
\label{lvertalt}\\
&&\hspace{5mm}
=~g(k_\perp|\calf^j\big( x_B+{k_\perp^2\over\alpha s}\big)\Big\{{\alpha  x_Bsg_{\mu i}-2k^\perp_{\mu}k_i\over\alpha x_Bs+k_\perp^2}(k_jU+Up_j){1\over\alpha  x_Bs+p_\perp^2}U^\dagger
\nonumber\\
&&\hspace{11mm}
-~2k^\perp_\mu U{ g_{ij}\over \alpha x_Bs+p_\perp^2}U^\dagger-~2g_{\mu j} U{p_i\over\alpha x_Bs+p_\perp^2}U^\dagger+{2k^\perp_\mu\over k_\perp^2}g_{ij}\Big\}|y_\perp)^{ab}   
~+~O(p_{2\mu})
\nonumber
\end{eqnarray}
where
the operator
$\calf_i(\beta)$
is defined as usual
\begin{eqnarray}
&&\hspace{-1mm}
(k_\perp|\calf_i(\beta)|y_\perp)
~\equiv~{2\over s}\int\! dy_\ast ~e^{i\beta y_\ast-i(k,y)_\perp}\calf_i(y_\ast, y_\perp)
\label{defoperf}
\end{eqnarray}
It is worth noting that at $ x_B=0$ this vertex agrees with the one obtained in Ref. \cite{bal04}.

The production part of the evolution kernel is proportional to the cross section of gluon emission given by the product 
of Eq. (\ref{lvertalt}) and complex conjugate vertex integrated according to Eq. (\ref{3.1}). To find the full kernel we should
add the virtual part.

\begin{figure}[htb]
\begin{center}
\includegraphics[width=141mm]{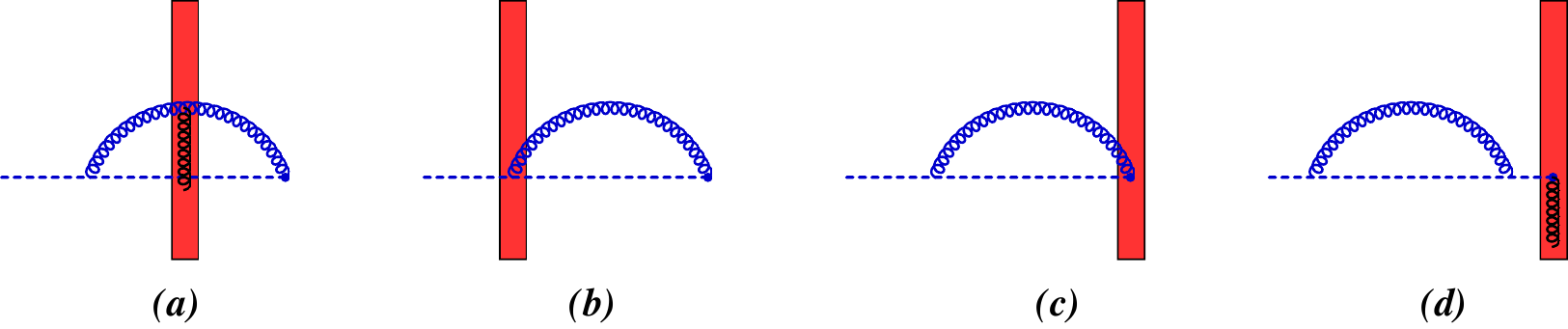}
\end{center}
\caption{Virtual part of the evolution kernel.\label{fig:5}}
\end{figure}

As in the case of production
kernel we  calculate the diagrams in Fig. \ref{fig:5}a, \ref{fig:5}b, and \ref{fig:5}c separately 
and then check that the final result does not depend on the size of the shock wave $\sigma_\ast$
(it is easy to see that the diagram in Fig. \ref{fig:5}d vanishes in Feynman gauge).
The result of the calculation is
\begin{eqnarray}
&&\hspace{-1mm}
\langle\calf_i^n( x_B, y_\perp)\rangle^{\rm Fig. ~5}~=~-ig^2f^{nkl}\!\int_{\sigma'}^{\sigma}\!{\dhd\alpha\over\alpha}
(y_\perp|-{p^j\over p_\perp^2}\calf_k( x_B)(i\!\stackrel{\leftarrow}{\partial}_l+U_l)\label{virtalt}\\
&&\hspace{31mm}
\times~(2\delta_j^k\delta_i^l-g_{ij}g^{kl})
U{1\over \alpha x_Bs+p_\perp^2}U^\dagger
+\calf_i( x_B){\alpha x_Bs\over p_\perp^2(\alpha x_Bs+p_\perp^2)}|y_\perp)^{kl}
\nonumber
\end{eqnarray}
where $\calf_k\stackrel{\leftarrow}{\partial}_l~\equiv~\partial_l\calf_k~=-i[p_l,\calf_k]$. For the complex conjugate amplitude one obtains
\begin{eqnarray}
&&\hspace{-1mm}
\langle \ticalf_i^n( x_B, x_\perp)\rangle^{\sigma} ~=~-ig^2f^{nkl}\!\int_{\sigma'}^{\sigma}{\!\dhd\alpha\over \alpha}
(x_\perp|
\tilU{1\over \alpha  x_Bs+p_\perp^2}\tilU^\dagger
\label{virtaltcc}\\
&&\hspace{25mm}
\times~(2\delta_i^k\delta_j^l-g_{ij}g^{kl} )(i\partial_k-\tilU_k)\tilde{\calf}_l( x_B)
{p^j\over p_\perp^2}
+~\tilcaf_i( x_B)
{\alpha x_Bs\over p_\perp^2(\alpha x_Bs+p_\perp^2)}|x_\perp)^{kl}
\nonumber
\end{eqnarray}
%

\section{Evolution equation for gluon TMD}

 Now we are in a position to assemble all leading-order contributions to the rapidity evolution of gluon TMD. Adding the production part 
 (\ref{3.1}) with Lipatov vertices (\ref{lvertalt}) and the virtual parts from previous Section (\ref{virtalt}) and (\ref{virtaltcc}) we obtain \cite{BT1}
\begin{eqnarray}
&&\hspace{-1mm}
{d\over d\eta}\langle p|\tilcaf_i^a( x_B, x_\perp) \calf_j^a( x_B, y_\perp)|p\rangle^{\eta=\ln\sigma}~~
\label{6.1}\\
&&\hspace{-1mm}
=~-\alpha_s\langle p|\!\int\!\dhd^2k_\perp
~{\rm Tr}\{\tiL_i^{~\mu}(k,x_\perp, x_B)^{\rm light-like}
L_{\mu j}(k,y_\perp, x_B)^{\rm light-like}\}
\nonumber\\
&&\hspace{-1mm}
+~2{\rm Tr}\Big\{\ticalf_i( x_B,x_\perp)
(y_\perp|-{p^m\over p_\perp^2}\calf_k( x_B)(i\!\stackrel{\leftarrow}{\partial}_l+U_l)(2\delta_m^k\delta_j^l-g_{jm}g^{kl})
U{1\over \alpha x_Bs+p_\perp^2}U^\dagger
\nonumber\\
&&\hspace{33mm}
+~\calf_j( x_B){\alpha x_Bs\over p^2_\perp(\alpha x_Bs+p_\perp^2)}|y_\perp)
\nonumber\\
&&\hspace{-1mm}
+~(x_\perp|
\tilU{1\over \alpha  x_Bs+p_\perp^2}\tilU^\dagger(2\delta_i^k\delta_m^l-g_{im}g^{kl} )(i\partial_k-\tilU_k)\ticalf_l( x_B)
{p^m\over p_\perp^2}
\nonumber\\
&&\hspace{33mm}
+~\tilcaf_i( x_B)
{\alpha x_Bs\over p_\perp^2(\alpha x_Bs+p_\perp^2)}|x_\perp)\calf_j\big( x_B, y_\perp\big)\Big\}|p\rangle~+~O(\alpha_s^2)
\nonumber
\end{eqnarray}
where Tr is a trace in the adjoint representation. 
\footnote{Here one can erase  tilde from Wilson lines since we have a sum over full set of states and gluon operators at space-like (or light-like) intervals commute with each other.}  
This equation describes the rapidity evolution of the operator (\ref{operator})  at any Bjorken $x_B$ and any transverse momenta. 

Let us discuss the gauge invariance of this equation. The l.h.s. is gauge invariant after taking into account gauge link at $+\infty$ as shown
in Eq. (\ref{inftylink}). As to the right side, it was obtained by calculation in the background field and promoting the background fields 
to operators in a usual way. However, we performed our calculations in a specific background field $A_\bu(x_\ast, x_\perp)$ with a finite support in $x_\perp$ and we need to address the question how can we restore the r.h.s. of Eq. (\ref{6.1}) in a generic field $A_\mu$. 
It is easy to see how one can restore the gauge-invariant form: just add gauge link at $+\infty p_1$ or $-\infty p_1$ appropriately. 
For example, the terms $U_z(z|{1\over\sigma\beta s+p_\perp^2}|z')U^\dagger_{z'}$ in r.h.s. of should be replaced by 
$U_z[z_\perp-\infty p_1,z'_\perp-\infty p_1](z|{1\over\sigma\beta s+p_\perp^2}|z')U^\dagger_{z'}$. After performing these insertions we will have the result which is (i) gauge invariant and (ii) coincides with Eq. (\ref{6.1}) for our choice of background field. At this step, the  background 
fields in the r.h.s. of Eq. (\ref{6.1}) can be promoted to operators. 

\section{Conclusions}

We have described the rapidity evolution of gluon TMD with Wilson lines going to $+\infty$ in the whole range of Bjorken $x_B$ and the whole range of transverse momentum $k_\perp$. It should be emphasized that with our definition of rapidity cutoff (\ref{cutoff}) the leading-order matrix elements 
of TMD operators are UV-finite so the rapidity evolution is the only evolution and it 
describes all the dynamics of gluon TMDs in the leading-log approximation.

As an outlook, it would be very interesting to obtain the NLO correction to the evolution equation (\ref{6.1}). The NLO corrections
to the BFKL  \cite{nlobfkl} and BK \cite{nlobk,kw,nlojimwlk} equation are available but they suffer from the well-known problem that they lead to negative cross sections. This difficulty can be overcome by the ``collinear resummation'' of double-logarithmic contributions for the BFKL
\cite{resumbfkl} and BK \cite{resumbk} equations and we hope that our Eq. (\ref{6.1}) and especially its future NLO version will 
help to solve the problem of negative cross sections of NLO amplitudes at high energies.

The authors are grateful to G.A. Chirilli, J.C. Collins, Yu. Kovchegov,  A. Prokudin,  A.V. Radyushkin, T. Rogers, and F. Yuan for valuable discussions. This work was supported by contract
 DE-AC05-06OR23177 under which the Jefferson Science Associates, LLC operate the Thomas Jefferson National Accelerator Facility, and by the grant DE-FG02-97ER41028.

\end{document}